\documentclass[twocolumn,showpacs]{revtex4}
\usepackage{graphics,epsfig,graphicx}
\usepackage{color}

\def\maj#1{\ifmmode\mbox{\usefont{U}{msb}{m}{n}#1}\else{\usefont{U}{msb}{m}{n}#1}\fi}
\def\v#1{\mathbf{#1}}

\begin{document}

\title{\textbf{BCS ansatz for superconductivity in the canonical ensemble\\ and the Pauli exclusion principle}}
\author{G. Zhu$^1$, M. Combescot$^{1,2}$ and O. Betbeder-Matibet$^2$}
\affiliation{(1)  Department of Physics, University of Illinois at Urbana-Champaign, 1110 W Green St, Urbana, IL, 61801}
\affiliation{(2) Institut des NanoSciences de Paris, Universit\'e Pierre et Marie Curie,
CNRS, Tour 22, 4 place Jussieu, 75005 Paris}


\begin{abstract}
The usual formulation of the BCS ansatz for superconductivity in the grand canonical ensemble makes the handling of the Pauli exclusion principle between paired electrons straightforward. It however tends to mask that many-body effects between Cooper pairs interacting through the reduced BCS potential are entirely controlled by this exclusion. To show it up, one has to work in the canonical ensemble. Pauli blocking between a fixed number of composite bosons is however known to be difficult to handle. To do it, we here develop a commutator formalism for Cooper pairs, along the line we used for excitons. We then rederive, within the $N$-pair subspace, a few results of BCS superconductivity commonly derived in the grand canonical ensemble, to evidence their Pauli blocking origin. We end by discussing what should be called ``Cooper pair wave function".
\end{abstract}


\date{\today}

\maketitle

\section{Introduction}

\begin{figure}[htbp]
\begin{center}
\includegraphics[width=0.4\textwidth]{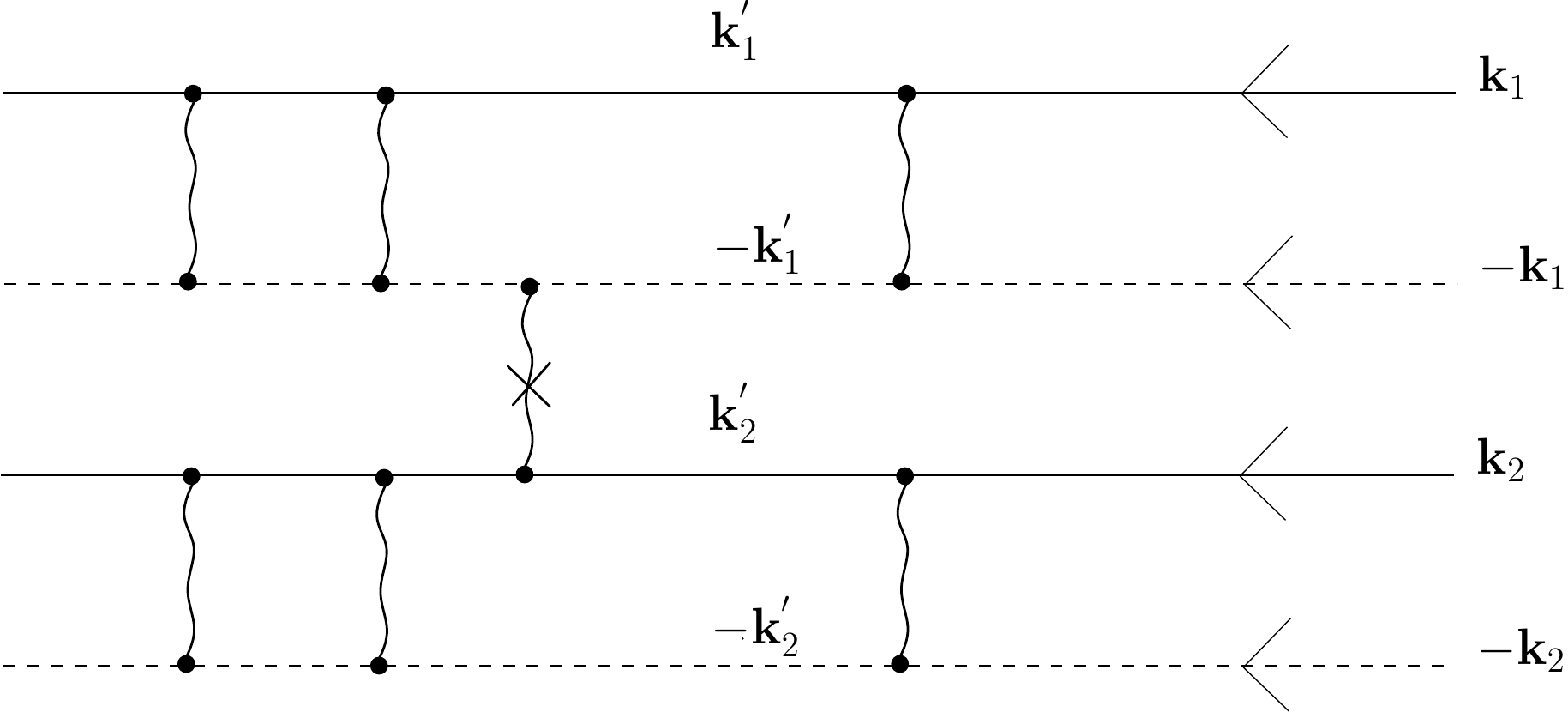}
\caption{Two Cooper pairs made of the repeated interactions of an up spin electron $\v k$ and a down spin electron $-\v{k}$ cannot interact by the reduced BCS potential because this would impose $\v{k}'_1=\v{k}'_2$.  The two up spin electrons would then have the same momentum which is impossible due to the Pauli exclusion principle. \label{fig:shiva}}

\end{center}
\end{figure}

A major breakthrough in the understanding of superconductivity is definitely due to the wave function ansatz proposed by Bardeen, Cooper, Schrieffer \cite{BCS}. The great advantage of this ansatz is that it leads to results in agreement with experiments through calculations quite easy to perform analytically. Its standard form in the grand canonical ensemble as a sum of  states with different particle numbers, makes the Pauli exclusion principle between paired up and down spin electrons straightforward to handle. Calculations in the grand canonical ensemble however mask the key role played by the Pauli exclusion principle in superconductivity. Indeed, due to the very peculiar form of the reduced BCS potential --- an up spin electron $(\v k)$ can interact with a down spin electron $(-\v k)$ only --- Pauli blocking is the only way two correlated electron pairs can feel each other (see Fig. \ref{fig:shiva}). This in particular explains why Cooper pairs can strongly overlap witho
 ut dissociating, by contrast with excitons which dissociate into an electron-hole plasma \cite{CN} through a Mott transition when overlap starts.
To better understand the key role played by the Pauli exclusion principle in the physics of BCS superconductors, it is necessary to stay in the canonical ensemble, with a pair number and a number of states available for pairing fixed. However, to exactly handle Pauli blocking between a fixed number of paired fermions like excitons or Cooper pairs, is far from easy. 
There were, in the past, several discussions on the exact particle-conserving solution within the reduced BCS potential using Richardson-Gaudin procedure\cite{Richardson1,Richardson2,Richardson3,Gaudin,CobosonBcsRich,ortiz,JETPLett}and its difference with the  BCS ansatz \cite{Bog,Bardeen,bang,hasegawa,Roman2002}.  Most discussions on the ground state however  focus on recovering the correct energy or some physical quantities, like the integrals of motion (\cite{ortiz}), instead of the wave function itself. Yet, the BCS ansatz wave function is strongly linked to the picture people commonly have of superconductivity. 

During the last decade, we developed  a many-body formalism appropriate to  composite bosons made of  fermion pairs\cite{CobosonPhysicsReports, combescotBCS, CobosonBcsRich,motheaten}. Up to now, we have extensively studied excitons and developed a formalism adapted to their many-body physics. Because the long-range Coulomb interaction between electrons and holes leads to a Mott dissociation of the exciton gas  into an electron-hole plasma \cite{CN} when the density increases, the relevant exciton regime is the dilute regime. By contrast, the relevant regime in BCS superconductivity is the dense regime with Cooper pair wave functions strongly overlapping. As a result, the many-body physics of composite bosons like Cooper pairs is expected to have some similarities with the one of excitons, but with a few major differences.

In section II, we  develop a commutator formalism for paired electrons capable to handle the Pauli exclusion principle within a $N$-pair condensate, in an exact way.

In section III, we come back to the BCS ansatz in its grand canonical form and we briefly rederive through the standard grand canonical procedure, some textbook results \cite{Tinkham,Fetter} on quantities we are going to consider in the canonical ensemble, namely, the $\v k$-state population, the pair number mean value, its fluctuations, and the two-pair correlation function.

In section IV, we turn to the canonical ensemble, with a fixed number of pairs. Through a direct study of the probability distribution for $N$-pair states in the BCS ansatz,  we prove that this ansatz indeed corresponds to a distribution very much peaked on a particular value $N^\ast$ of the pair number, as a result of the ``moth-eaten effect'' induced on Cooper pairs by the Pauli exclusion principle\cite{motheaten}. The standard derivation of this result, through the fluctuations of the pair number around its mean value\cite{Tinkham}, completely hides the microscopic origin of this maximum.

In section V, we calculate the fraction of the $\v k$ electron state occupied in a $N$-pair state. We do
show that it is identical to the one calculated within the grand canonical version of the ansatz for $N=N^\ast$. This is also true for the pair operator mean value associated to what is often called ``pair wave function''.

In section VI, we come back to what should be called ``Cooper pair wave function''.

In section VII, we conclude.

\section{Composite boson formalism for condensed pairs}

The goal of this section is to develop a formalism capable to, in an exact way, handle Pauli blocking within a $N$-fermion-pair condensate. 

For that, we follow our previous works  \cite{ CobosonBcsRich,motheaten} and introduce the generalized creation operator for correlated pairs,
\begin{equation}
B_n^\dag=\sum_{\v k}|\varphi_{\v k}^2|^n\varphi_{\v k}\beta_{\v k}^\dag\ ,
\end{equation}
with $n=(0,1,2\cdots)$. The operator $\beta_{\v k}^\dag=a_{\v k}^\dag b_{-\v k}^\dag$ creates a pair of free fermions with zero total momentum. In the case of BCS superconductivity, these fermions are up and down spin electrons.

We first note that free fermion pair creation operators commute, $[\beta_{\v k'}^\dag,\beta_{\v k}^\dag]=0$, while
\begin{equation}
[\beta_{\v k'},\beta_{\v k}^\dag]=\delta_{\v k',\v k}-D_{\v k'\v k}\ ,
\end{equation}
with $D_{\v k'\v k}=\delta_{\v k',\v k}(a_{\v k}^\dag a_{\v k}+b_{-\v k}^\dag b_{-\v k})$, so that $D_{\v k'\v k}|0\rangle=0$. We also have
\begin{equation}
[a_{\v p}^\dag a_{\v p},\beta_{\v k}^\dag]=\delta_{\v p,\v k}\beta_{\v k}^\dag=[b_{-\v p}^\dag b_{-\v p},
\beta_{\v k}^\dag]\ .
\end{equation}

It is then easy to show that
\begin{equation}\label{BBcomm}
[B_m,B_n^\dag]=\tau_{m+n}-D_{m+n}\ ,
\end{equation}
where the ``deviation from boson operator'' of these generalized correlated pairs is given by $D_m=\sum_{\v k}|\varphi_{\v k}^2|^{m+1}(a_{\v k}^\dag a_{\v k}+b_{-\v k}^\dag b_{-\v k})$. The scalar $\tau_m$, defined as
\begin{equation}
\tau_m=\sum_{\v k}|\varphi_{\v k}^2|^{m+1},
\end{equation}
is the $m+1$ moment of the $\v k$ state distribution in the correlated pair at hand. To possibly relate this moment to the correlated pair wave function, we are led to normalize the $\varphi_{\v k}$ distribution as $\tau_0=\sum_{\v k}|\varphi_{\v k}^2|=1$. We then have, for $|0\rangle$ being the vacuum state
\begin{equation}
\langle0|B_0^{}B_0^{\dag}|0\rangle=\tau_0=1
\end{equation}

In order to easily handle the Pauli exclusion principle within a $B_0^{\dag N}|0\rangle$ condensate, we also need
\begin{equation}
[D_m,B_n^\dag]=2B_{m+n+1}^\dag\ .
\end{equation}
Using it and
\begin{eqnarray}
\left[D_m,B_0^{\dag N}\right]=\left[D_m,B_0^\dag\right]B_0^{\dag N-1}\hspace{1cm}\nonumber\\
+B_0^\dag\left[D_m,B_0^{\dag N-1}\right],
\end{eqnarray}
we get by iteration
\begin{equation}
\left[D_m,B_0^{\dag N}\right]=2NB_{m+1}^\dag B_0^{\dag N-1}\ .\label{DBcomm}
\end{equation}
In the same way, Eqs.(\ref{BBcomm}) and (\ref{DBcomm}), along with
\begin{eqnarray}
\left[B_m,B_0^{\dag N}\right]=\left[B_m,B_0^\dag\right]B_0^{\dag N-1}\hspace{2cm}\nonumber\\
+B_0^\dag\left[B_m,B_0^{\dag N-1}\right],\label{Bcomm}
\end{eqnarray}
allow us to rewrite the RHS of the above equation as
\begin{equation}\label{BcommRHS}
NB_0^{\dag N-1}(\tau_m-D_m)-N(N-1)B_{m+1}^\dag B_0^{\dag N-2}.
\end{equation}

One important quantity for a condensate made of $N$ composite bosons $B_0^\dag$ is its normalization factor. Let us write it as
\begin{equation}
\langle 0|B_0^NB_0^{\dag N}|0\rangle=N!F_N\ .
\end{equation}
If $B_0^\dag$ were an elementary boson creation operator, we would have $F_N=1$. For composite bosons, $F_N$, equal to 1 for $N=1$, decreases when $N$ increases, due to what we called ``moth eaten effect''\cite{JETPLett}: more and more free pair states are missing in the $B_0^\dag$ operators of $B_0^{\dag N}|0\rangle$ due to the Pauli exclusion principle between these $N$ pairs as if $N$ little moths had eaten these free states. 

To calculate $F_N$, we first note that 
$\langle 0|{B_0^N}B_0^{\dag}{}^{ N}|0\rangle$ also reads $\langle 0|B_0^{N-1}B_0B_0^{\dag}{}^ N|0\rangle$. We then use Eqs. (\ref{Bcomm}, \ref{BcommRHS}). For $\tau_0=1$, we find 
\begin{equation}\label{FN}
F_N=F_{N-1}-\frac{1}{(N-2)!}\langle\langle 0|B_0^{N-1}B_1^\dag B_0^{\dag N-2}|0\rangle,
\end{equation}
and we iterate using Eqs.(\ref{Bcomm}, \ref{BcommRHS}). This shows that the $F_N$'s are linked by
\begin{eqnarray}
F_N&=&F_{N-1}-(N-1)\tau_1F_{N-2}\hspace{2.3cm}\nonumber\\
&&\hspace{0.5cm}+(N-1)(N-2)\tau_2F_{N-3}+\cdots\nonumber\\
&&\hspace{1cm}+(-1)^{N-1}(N-1)!\tau_{N-1}F_0
\end{eqnarray}
Eq.(\ref{FN}) also shows that $F_N$ is a decreasing function of $N$:  The moth-eaten effect gets larger and larger when $N$ increases. Indeed, the last matrix element is positive as seen by expanding it on free pair operators; this matrix element then reads
\begin{equation}
\sum_{\v k_1\cdots \v k_{N-1}}^{\neq}|\varphi_{\v k_1}^4||\varphi_{\v k_2}^2|\cdots|\varphi_{\v k_{N-1}}^2|,
\end{equation}
the sum being taken over different $(\v k_1,\cdots,\v k_{N-1})$ due to the Pauli exclusion principle.

\section{BCS ansatz}

Let us introduce the \emph{unnormalized} correlated pair creation operator
\begin{equation}\label{C}
C^\dag=\sum_{\v p}\phi_{\v p}\beta_{\v p}^\dag\ .
\end{equation}
with $\langle 0|CC^{\dag }|0\rangle$ possibly different from $1$. From it,  we construct the following linear combination of $N$-pair states\cite{Tinkham,Fetter,Leggett}
\begin{equation}
|\phi\rangle=\sum_{N=1}^{+\infty}\frac{1}{N!}C^{\dag N}|0\rangle
\end{equation}
This state also reads
\begin{equation}
|\phi\rangle=e^{C^\dag}|0\rangle=\Pi_{\v p}(1+\phi_{\v p}\beta_{\v p}^\dag)|0\rangle.
\end{equation}
since $\beta_{\v p}^{\dag 2}=0$ due to the Pauli exclusion principle. By writing $\phi_{\v p}$ as $v_{\v p}/u_{\v p}$ with $|u_{\v p}^2|+|v_{\v p}^2|=1$, we get the usual form of the \emph{normalized} BCS state as product of $\v p$ operators
\begin{equation}
|\phi_{BCS}\rangle=\gamma|\phi\rangle
=\prod_{\v p}(u_{\v p}+v_{\v p}\beta_{\v p}^\dag)|0\rangle,
\end{equation}
where the normalization factor reads $\gamma=\prod_{\v p}u_{\v p}$.

Using this $\v p$ product, the $\v k$ electron distribution in the $|\phi_{BCS}\rangle$ state is easy to find as
\begin{eqnarray}
\langle\hat{N}_{\v k}\rangle&=&\langle\phi_{BCS}|a_{\v k\uparrow}^\dag a_{\v k\uparrow}|\phi_{BCS}\rangle\nonumber\\
&=&|v_{\v k}^2|=\frac{|\phi_{\v k}^2|}{1+|\phi_{\v k}^2|}\ .\label{Nk}
\end{eqnarray}
As a result, the $\phi_{\v k}$ distribution of the correlated pair operator $C^\dag$ defined in Eq.(\ref{C}) is related to the mean value $\langle\hat{N}\rangle$ of the number of up \emph{or} down spin electrons $\hat{N}=\sum_{\v k}a_{\v k\uparrow}^\dag a_{\v k\uparrow}$ in the $|\phi_{BCS}\rangle$ state through
\begin{eqnarray}
\langle\hat{N}\rangle&=&\sum_{\v k}\langle\hat{N}_{\v k}\rangle=\sum_{\v{k}}|v_{\v k}^2|
=\sum_{\v k}\frac{|\phi_{\v k}^2|}{1+|\phi_{\v k}^2|}\ .
\end{eqnarray}

Turning to the fluctuation of this mean value, we find that it reads
\begin{equation}
\frac{\langle\hat{N}^2\rangle-\langle\hat{N}\rangle^2}{\langle\hat{N}\rangle^2}=\frac{\sum_{\v k}|u_{\v k}^2v_{\v k}^2|}{\Big{(}\sum_{\v k}|v_{\v k}^2|\Big{)}^2}.
\end{equation}
Since each sum over $\v k$ is proportional to the sample volume, the above ratio goes to zero in the large sample limit as 1 over the volume. So, this fluctuation is indeed very small which proves that the $N$ distribution in the $|\phi_{BCS}\rangle$ state is very much peaked on its mean value $\langle\hat{N}\rangle$.

We can also consider the mean value of the two-pair operator $\beta_{\v k}\beta_{\v k'}^\dag$ in the $|\phi_{BCS}\rangle$ state. For $\v k\neq\v k'$, we find
\begin{equation}\label{betaBCS}
\langle\phi_{BCS}|\beta_{\v k}\beta_{\v k'}^\dag|\phi_{BCS}\rangle=u_{\v k}^\ast v_{\v k}u_{\v k'}v_{\v k'}^\ast=F_{\v k}F_{\v k'}^\ast\ 
\end{equation}
where $F_{\v k}$, often called ``pair wave function'', is defined as
\begin{equation}\label{fn}
F_{\v k}=u_{\v k}^\ast v_{\v k}=\frac{\phi_{\v k}}{1+|\phi_{\v k}^2|}\ .
\end{equation}

\section{Direct calculation of the $N$-pair state probability in the BCS ansatz}

The above calculation of the $\langle\hat{N}\rangle$ fluctuation using the grand canonical form $|\phi_{BCS}\rangle$ of the BCS ansatz as a $\v p$ product is quite convincing to conclude that the $N$-pair state distribution in this ansatz is very much peaked on the $N$ mean value. However, a direct calculation of the probability distribution, using the $N$ sum $|\phi\rangle$, is of interest to understand the microscopic origin of this peaked value.

If we only consider  the $(1/N!)$ prefactor in the $|\phi\rangle$ sum, we could na\"{\i}vely conclude that the $N=1$ state dominates the sum. However, in order for the prefactors of the $N$-pair state in the $|\phi\rangle$ expansion to have some physical meaning, this $N$-pair state must be normalized.

 To do it, we first introduce the normalized correlated pair operator
\begin{equation}
B_0^\dag=\sum_{\v p}\varphi_{\v p}\beta_{\v p}^\dag=\frac{C^\dag}{\sqrt{S}}\ ,
\end{equation}
where $\varphi_{\v p}=\phi_{\v p}/\sqrt{S}$ and $S=\sum_{\v p}|\phi_{\v p}^2|$, in order to have $\langle 0|B_0B_0^\dag|0\rangle=1$. The normalized $N$-pair state associated to the $C^\dag$ correlated pair operator then reads
\begin{equation}\label{psiN}
|\psi_N\rangle=\frac{B_0^{\dag N}|0\rangle}{\sqrt{N!F_N}}=\frac{C^{\dag N}|0\rangle}{\sqrt{N!F_NS^N}}\ 
\end{equation}
with $F_N$ defined as in Eq.(12), in order to have $\langle\psi_N|\psi_N\rangle=1$.

The above equation allows us to rewrite $|\phi_{BCS}\rangle$ given in Eq.(19) as
\begin{equation}\label{phiBCS}
|\phi_{BCS}\rangle=\gamma\sum_{N}\sqrt{\frac{F_NS^N}{N!}}|\psi_N\rangle
\equiv\sum_{N}\lambda_N|\psi_N\rangle\ .
\end{equation}
We then note that $\sum_N |\lambda_N^2|=1$ since $\langle\phi_{BCS}|\phi_{BCS}\rangle=1$ and $\langle\psi_N|\psi_N\rangle=1$; so, $|\lambda_N^2|$ indeed is the probability distribution of the $N$-pair states in the BCS ansatz.

To show that this $N$ distribution is peaked, we first note that $1/N!$ and $F_N$ both decrease from $1$ when $N$ increases. So, the ratio $F_N/N!$ also decreases when $N$ increases. In order to show that this decrease is compensated by the increase of $S^N$ in order for $\lambda_N$ to be possibly peaked, we first note that the sum $S=\sum_{\v p}|v_{\v p}^2|/|u_{\v p}^2|$ is  larger than $\sum_{\v p}|v_{\v p}^2|=\langle\hat{N}\rangle$ since $|u_{\v p}^2|+|v_{\v p}^2=1$ so $|u_{\v p}^2|$ is smaller than 1. As a result, $S> 1$ and $S^N$ increases with $N$. This $S^N$ increase first dominates the $1/N!$ decrease  since, due to the Stirling formula, $S^N/N!\simeq(S/N)^N$. So, in the absence of the $F_N$ factor, the $\lambda_N$ probability distribution distribution would be peaked on $N^{\ast\ast}\simeq S$, which is far larger than the pair number mean value $\langle\hat{N}\rangle$.

The moth-eaten effect induced by Pauli blocking on Cooper pairs is responsible for bringing the $\lambda_N$ peak back to $\langle\hat{N}\rangle$, through the $F_N$ decrease it induces.  To show it, we first note that this $F_N$ decrease does not affect the $\lambda_N$ behavior so much for small $N$ since, as seen from Eq.(\ref{FN}), $F_N/F_{N-1}$ stays close to 1 for  small $N$. By contrast, $F_N$ is going to play a key role for large $N$'s by changing the peak of the probability distribution from $N^{\ast\ast}=S$ to $N^\ast$ which, as a definition of the $\lambda_N$ peak, must be such that 
$\lambda_{N^\ast-1}\simeq \lambda_{N^\ast}$. Eq.(\ref{phiBCS}) then gives
\begin{equation}\label{xn}
1\simeq\frac{F_{N^\ast-1}}{F_{N^\ast}}\ \frac{N^\ast}{S}\equiv x_{N^\ast}^2\ 
\end{equation}
If Cooper pairs were elementary bosons, $F_N$ would stay equal to 1 for all $N$ and the peak of the $\lambda_N$ distribution would take place for a pair number equal to $S$ which is far larger than  $\langle\hat{N}\rangle$. For composite bosons, the $F_N/F_{N-1}$ ratio is smaller than 1 since $F_N$ decreases with $N$, as previously shown; so, $\lambda_N$ is peaked on a $N$ value smaller than S.

If the $N^\ast$ peak were in the dilute regime, the $F_N$ ratio would be close to 1 and $N^\ast$ would be close to $S\gg\langle\hat{N}\rangle$. This shows that the solution of Eq.(\ref{xn}) is in the dense regime, with $F_N/F_{N-1}$ substantially smaller than 1. This has to be contrasted to excitons for which the $F_N/F_{N-1}$ ratios always are close to 1: at large density, excitons would dissociate into an electron-hole plasma\cite{CN}.

More demanding is to go further and to precisely relate the pair number $N^*$ corresponding to the $\lambda_N$ maximum to the pair number mean value  $\langle\hat{N}\rangle$ calculated within the $\v p$ product form of the BCS ansatz. With this goal in mind, let us first calculate the $\v k$ electron state population in the $B_0^{\dag N}$ condensate.

\section{$\v k$-electron population in $N$-pair state}

The $\v k$ electron distribution is straightforward to obtain in the $|\phi_{BCS}\rangle$ state; the result is given in Eq.(\ref{Nk}). To calculate the $\v k$ electron distribution in a $N$-pair state $|\psi_N\rangle$ is not as easy.

As a first idea, we could perform a brute force calculation of
\begin{equation}\label{NkK}
 \langle\hat{N}_{\v{k}}\rangle_{N}=\langle\psi_N|a_{\v k\uparrow}^\dag a_{\v k\uparrow}|\psi_N\rangle
\end{equation}
in the normalized state $|\psi_N\rangle$ given in Eq.(\ref{psiN}), using the commutator formalism developed in section II. To do it, we start with $[a_{\v k},B_0^\dag]=\varphi_{\v k}b_{-\v k}^\dag$ which gives by iteration
\begin{equation}
[a_{\v k},B_0^{\dag N}]=N\varphi_{\v k} b_{-\v k}^\dag B_0^{\dag N-1}\ .
\end{equation}
This allows us to rewrite Eq.(\ref{NkK}) as
\begin{equation}
 \langle\hat{N}_{\v{k}}\rangle_{N}=\frac{N^2|\varphi_{\v k}^2|}{N!F_N}\langle 0|B_0^{N-1}b_{-\v k}b_{-\v k}^\dag B_0^{\dag N-1}|0\rangle\ .
\end{equation}
We then use $b_{-\v k}b_{-\v k}^\dag=1-b_{-\v k}^\dag b_{-\v k}$ and iterate the process. This gives $ \langle\hat{N}_{\v{k}}\rangle_{N}$ through the following $F_N$ expansion
\begin{eqnarray}
 \langle\hat{N}_{\v{k}}\rangle_{N}=\frac{N}{F_N}\Big{[}|\varphi_{\v k}^2|F_{N-1}-(N-1)|\varphi_{\v k}^4|F_{N-2}
\nonumber\\
+(N-1)(N-2)|\varphi_{\v k}^6|F_{N-3}\cdots\Big{]}.
\end{eqnarray}
Eq.(14) shows that the sum over $\v k$ of the above bracket reduces to $F_N$; so we do have
\begin{equation}
\sum_{\v k} \langle\hat{N}_{\v{k}}\rangle_{N}=N,
\end{equation}
 as expected. However, through this $F_N$ expansion, it is not easy to show that $ \langle\hat{N}_{\v{k}}\rangle_{N}$ indeed reduces to $ \langle\hat{N}_{\v{k}}\rangle$ for $N$ equal to the peak value $N^\ast$.

A better way to make such a link is to follow Leggett \cite{Leggett} and to introduce the operator $C^\dag$ defined in Eq.(\ref{C}) with the $\v k$ state removed from the sum, namely,
\begin{equation}
C_{\v k}^\dag=\sum_{\v p\neq\v k}\phi_{\v p}\beta_{\v p}^\dag.
\end{equation}
We then construct its normalized form $B_{\v k}^\dag=C_{\v k}^\dag/\sqrt{S_{\v k}}$ with $S_{\v k}=\sum_{\v p\neq\v k}|\phi_{\v p}^2|$ in order to have $\langle 0|B_{\v k}B_{\v k}^\dag|0\rangle=1$. The associated $N$-pair normalized state then reads
\begin{equation}
|\psi_{N,\v k}\rangle=\frac{B_{\v k}^{\dag N}|0\rangle}{\sqrt{N!F_{N,\v k}}}=
\frac{C_{\v k}^{\dag N}|0\rangle}{\sqrt{N!F_{N,\v k}S_{\v k}^N}}\ ,
\end{equation}
where, as for $F_N$, the normalization factor $F_{N,\v k}$ is defined as $\langle 0|B_{\v k}^NB_{\v k}^{\dag N}|0\rangle=N!F_{N,\v k}$ in order to have $\langle\psi_{N,\v k}|\psi_{N,\v k}\rangle=1$.

By writing $C^{\dag N}$ as $(C_{\v k}^\dag+\phi_{\v k}\beta_{\v k}^\dag)^N$ and by noting that $(\beta_{\v k}^\dag)^2=0$ due to the Pauli exclusion principle, we easily find that the normalized $N$-pair state $|\psi_N\rangle$ defined in Eq.(\ref{psiN}) also reads
\begin{equation}
|\psi_N\rangle=\frac{|\psi_{N,\v k}\rangle+x_{N,\v k}\phi_{\v k}\beta_{\v k}^\dag|\psi_{N-1,\v k}\rangle}{\sqrt{1+x_{N,\v k}^2
|\phi_{\v k}^2|}}\ ,
\end{equation}
with  $x_{N,\v k}$  defined as $x_{N}$ in Eq. (\ref{xn}), namely
\begin{equation}
x_{N,\v k}^2=\frac{F_{N-1,\v k}}{F_{N,\v k}}\,\frac{N}{S_{\v k}}\ .
\end{equation}
 
Using this new expression of $|\psi_N\rangle$, it becomes easy to write the population of the $\v k$ electron state in the $N$ pair condensate in a compact form as
\begin{equation}
 \langle\hat{N}_{\v{k}}\rangle_{N}=\frac{x_{N,\v k}^2|\phi_{\v k}^2|}{1+x_{N,\v k}^2|\phi_{\v k}^2|}\ .
\end{equation}
When compared to $ \langle\hat{N}_{\v{k}}\rangle$ calculated in the $|\phi_{BCS}\rangle$ state, as given in Eq.(\ref{Nk}), we see that $ \langle\hat{N}_{\v{k}}\rangle_{N}$ calculated in a $N$-pair condensate reduces to $ \langle\hat{N}_{\v{k}}\rangle$ for $x_{N,\v k}^2=1$. We then note that, for large samples, the number of $\v k$ states is very large; so, the $S$ sum does not change very much when one state is removed, i.e., $S\simeq S_{\v k}$. In the same way $F_N\simeq F_{N,\v k}$. As a result, the condition $x_{N,\v k}^2=1$ also reads $1\simeq x_N^2=NF_{N-1}/SF_N$. This is fulfilled for $N=N^\ast$ defined in Eq.(\ref{xn}). This shows that  $\langle\hat{N}_{\v{k}}\rangle_{N^\ast}= \langle\hat{N}_{\v{k}}\rangle$: as expected, the $\v k$-electron state population calculated in a  $N$-pair state is equal to the one calculated in the grand canonical state $|\phi_{BCS}\rangle$ provided that $N$ corresponds to the maximum value $N^\ast$ of the $N$-pair distribution in the BCS st
 ate.

To get a better understanding of the link which exists between $|\phi_{BCS}\rangle$ and its $N$-pair state projection $|\Psi_N\rangle$, we can also calculate the mean value of the two-pair operator $\beta_{\v k}^{}\beta_{\v k'}^\dag$ in this $|\Psi_N\rangle$ state. As for $a_{\v k\uparrow}^\dag a_{\v k\uparrow}$, a brute force calculation of this mean value would give it as a $F_N$ expansion, not easy to compare with Eqs.(\ref{betaBCS}, \ref{fn}). We can instead follow Leggett \cite{Leggett} and introduce the operator $C^\dag$ in which the $\v k$ and $\v k'$ states are missing, namely,
\begin{equation}
C_{\v k\v k'}^\dag=\sum_{\v p\neq (\v k,\v k')}\phi_{\v p}\beta_{\v p}^\dag\ .
\end{equation}
We again construct its normalized form $B_{\v k\v k'}^\dag=C_{\v k\v k'}^\dag/\sqrt{S_{\v k\v k'}}$ where $S_{\v k\v k'}=\sum_{\v p\neq (\v k,\v k')}|\phi_{\v p}^2|$ and the associated $N$-pair normalized state
\begin{equation}
|\psi_{N,\v k\v k'}\rangle=\frac{B_{\v k\v k'}^{\dag N}|0\rangle}{\sqrt{N!F_{N,\v k\v k'}}}=
\frac{C_{\v k\v k'}^{\dag N}|0\rangle}{\sqrt{N!F_{N,\v k\v k'}S_{\v k\v k'}^N}},
\end{equation}
with $F_{N,\v k\v k'}$ such that $\langle 0|B_{\v k\v k'}^NB_{\v k\v k'}^{\dag N}|0\rangle=N!F_{N,\v k\v k'}$.

From $C^\dag=C_{\v k\v k'}^\dag+\phi_{\v k}\beta_{\v k}^\dag+\phi_{\v k'}\beta_{\v k'}^\dag$, it is then easy to show that the normalized $N$-pair state $|\psi_N\rangle$ defined in Eq.(\ref{psiN}) also reads
\begin{eqnarray}
|\psi_N\rangle=\frac{1}{\mathcal{N}}\Big{[}|\psi_{N,\v k\v k'}\rangle\hspace{4cm}\nonumber\\
+x_{N,\v k\v k'}(\phi_{\v k}\beta_{\v k}^\dag+
\phi_{\v k'}\beta_{\v k'}^\dag)|\psi_{N-1,\v k\v k'}\rangle\hspace{1cm}\nonumber\\
+x_{N,\v k\v k'}x_{N-1,\v k\v k'}\phi_{\v k}\phi_{\v k'}\beta_{\v k}^\dag\beta_{\v k'}^\dag|\psi_{N-2,\v k\v k'}\rangle\Big{]},
\label{psiNexpand}
\end{eqnarray}
where $x_{N,\v k\v k'}$ has a similar form as $x_{N,\v k}$ with now two states excluded instead of one, namely,
\begin{equation}
x_{N,\v k\v k'}^2=\frac{F_{N-1,\v k\v k'}}{F_{N,\v k\v k'}}\,\frac{N}{S_{\v k\v k'}}\ .
\end{equation}
In order to still have $\langle\psi_N|\psi_N\rangle=1$, the normalization factor $\mathcal{N}$ in Eq.(\ref{psiNexpand}) must be equal to
\begin{eqnarray}
\mathcal{N}=\Big{[}1+x_{N,\v k\v k'}^2\left(|\phi_{\v k}^2|+|\phi_{\v k'}^2|\right)\hspace{2cm}\nonumber\\
+x_{N,\v k\v k'}^2x_{N-1,\v k\v k'}^2|\phi_{\v k}^2||\phi_{\v k'}^2|\Big{]}^{1/2}\ .
\end{eqnarray}
If we now note that, for $N$ large, $x_{N,\v k\v k'}$ and $x_{N-1,\v k\v k'}$ are very close, this normalization factor reduces to a product
$\Big{[}1+x_{N,\v k\v k'}|\phi_{\v k}^2|\Big{]}^{1/2}\Big{[}1+x_{N,\v k\v k'}|\phi_{\v k'}^2|\Big{]}^{1/2}$. It is then easy to show, using Eq.(41) for $|\psi_N\rangle$, that
\begin{eqnarray}
\langle\psi_N|\beta_{\v k}\beta_{\v k'}^\dag|\psi_N\rangle=\hspace{3cm}\nonumber\\
\frac{x_{N,\v k\v k'}\phi_{\v k}}{1+x_{N,\v k\v k'}^2|\phi_{\v k}^2|}\ 
\frac{x_{N,\v k\v k'}\phi_{\v k'}}{1+x_{N,\v k\v k'}^2|\phi_{\v k'}^2|}\ .
\end{eqnarray}

When compared to the mean value of $\beta_{\v k}^{}\beta_{\v k'}^\dag$  calculated in the $|\phi_{BCS}\rangle$ state, as given in Eqs.(\ref{betaBCS}, \ref{fn}), we  find that these two results are identical provided that $1=x_{N,\v k\v k'}^2$. We then note that $x_{N,\v k\v k'}^2$ and $ x_N^2$  again are very close because the missing $(\v k , \v k')$ states do not very much affect $F_N$ and the $S$ sum in the large sample limite when the number of $\v k$'s is very large; so $F_{N,\v k\v k'}\simeq F_N$ and $S_{\v k\v k'}\simeq S$. We thus end with the same $\beta_{\v k}^{}\beta_{\v k'}^\dag$ mean values in the $N$ pair state $|\psi_N\rangle$ and in the $|\phi_{BCS}\rangle$ state provided that $N$ is equal to the maximum $N^\ast$ of the $\lambda_N$ distribution of $N$-pair states in the BCS ansatz, which also is the mean value of the particle number  $\langle\hat{N}\rangle$ in the $|\phi_{BCS}\rangle$ state.

\section{Cooper pair wave function}

We wish to end this work by reconsidering what should be called ``pair-wave function'', since $F_{\v{k}}$ defined in Eq. (\ref{fn}) is often called this way.
Cooper pairs like excitons are composite bosons made of two fermions. Their creation operators thus read as a sum of free fermion pair creation operators $\beta_{\v k}^\dag$. To properly identify what should be called ``Cooper pair wave function'', it is enlightening to compare their creation operator with the exciton creation operator.

We commonly distinguish two types of excitons: Wannier excitons\cite{Wannier} and Frenkel excitons\cite{Frenkel,frenkel}, the latter having closer similarity with Cooper pairs. Indeed, Wannier excitons are constructed on free electrons and free holes, so that they are ``double index'' composite bosons. Their creation operators read
\begin{equation}\label{Bi}
B_i^\dag=\sum_{\v k_e,\v k_h}a_{\v k_e}^\dag b_{\v k_h}^\dag \langle\v k_h,\v k_e|i\rangle\ ,
\end{equation}
with $i=(\v Q_i,\nu_i)$, where $\v Q_i$ is the exciton center-of-mass momentum and $\nu_i$ its relative motion index. The prefactor $\langle\v k_h,\v k_e|i\rangle$ of the $i$ exciton expansion on free electron-hole pairs, is the $i$ exciton wave function in momentum space.

By contrast, Frenkel excitons are made of atomic excitations for atoms on a regular lattice. They thus are ``single index'' composite bosons, their creation operators reading as
\begin{equation}
B_{\v Q}^\dag=\sum_n\frac{e^{i\v Q.\v R_n}}{\sqrt{N_s}} a_n^\dag b_n^\dag.
\end{equation} 
$\v R_n$ is the position of the excited atom and $N_s$ the number of atomic sites.

Cooper pairs also are ``single index'' composite bosons since an up spin electron $\v k$ is coupled to one down spin electron $(-\v k)$ only by the reduced BCS potential. The (normalized) creation operator of one Cooper pair in the BCS condensate reads, as quoted above,
\begin{equation}\label{B0}
B_0^\dag=\sum_{\v k}\varphi_{\v k}a_{\v k}^\dag b_{-\v k}^\dag,
\end{equation}
where $\varphi_{\v k}=\phi_{\v k}/\sqrt{S}$ with $\phi_{\v k}=v_{\v k}/u_{\v k}$ and $S=\sum_{\v k}|\phi_{\v k}^2|$.

In view of Eqs.(\ref{Bi}-\ref{B0}), $\varphi_{\v k}$ must be taken as the Cooper pair wave function, without any ambiguity, in spite of other quantities like $F_{\v k}=u_{\v k}^\ast v_{\v k}$ often quoted as ``pair wave function" in the literature. 

\begin{figure}[htbp]
\begin{center}
\includegraphics[width=0.3\textwidth]{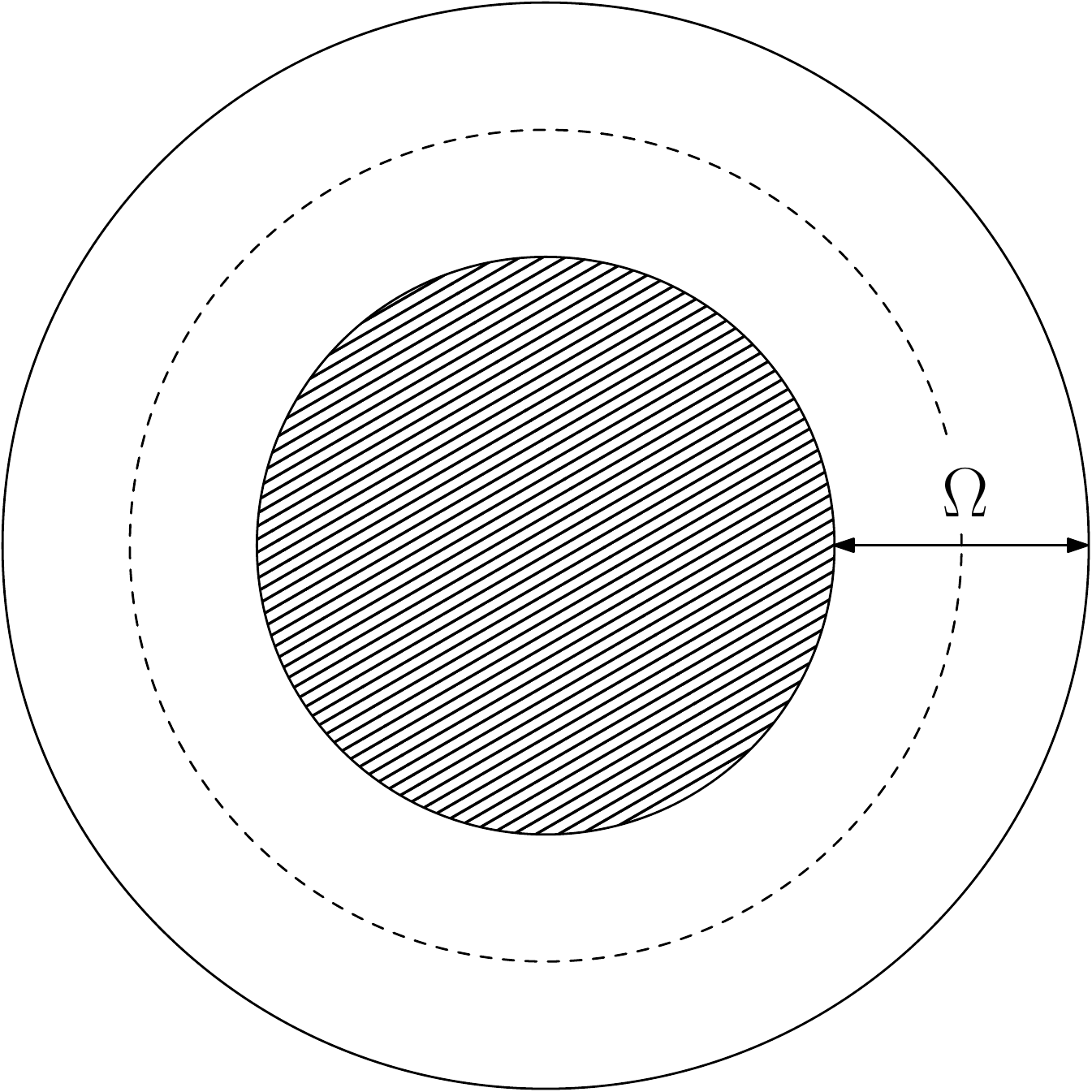}
\caption{The reduced BCS potential extends over an energy $\Omega$ of the order of twice the phonon energy, above a non-interacting electron Fermi sea with Fermi energy $\epsilon_{F_0}$.  The dashed line indicates the chemical potential $\mu$.  \label{potfig}} 

\end{center}
\end{figure}

Although Cooper pairs and Frenkel excitons both are ``single index'' composite bosons, they however have a few major differences. One of them comes from the fact that the Frenkel exciton wave function is just a phase, so that the distribution of the  excited site $n$  in a Frenkel exciton is flat. By contrast, the Cooper pair wave function is given by $\varphi_{\v k}=\phi_{\v k}/\sqrt{S}$ with
\begin{equation}\label{phi}
\phi_{\v k}=\frac{v_{\v k}}{u_{\v k}}=\sqrt{\frac{E_{\v k}-\xi_{\v k}}{E_{\v k}+\xi_{\v k}}}=\frac{\Delta}{E_{\v k}+\xi_{\v k}}
\end{equation}
 $\xi_{\v k}=\epsilon_{\v k}-\mu$ and $E_{\v k}^2=\xi_{\v k}^2+\Delta^2$, the electron $\v k$ energy being $\epsilon_{\v k}=\v k^2/2m$ while $\mu$ is the chemical potential of the $|\phi_{BCS}\rangle$ state in the grand canonical ensemble. In the usual BCS configuration with a reduced BCS potential extending symmetrically on a phonon energy scale on both sides of the normal electron Fermi sea, $\mu$ is in the middle of the layer over which the potential acts, this layer extension $\Omega$ being twice the phonon energy (see Fig. \ref{potfig}). The gap $\Delta$ then reads in the small potential limit as $\Delta\simeq \Omega e^{-1/\rho_0V}$, where $\rho_0$ is the density of states taken as constant over the layer where the potential acts. As a result, $\phi_{\v k}$ has three different scales: $\phi_{\v k}\simeq 1$ for $\epsilon_{\v k}$ very close to $\mu$ on the $\Delta$ scale. $\phi_{\v k}\simeq 2e^{1/\rho_0V}(\mu-\epsilon_{\v k})/\Omega$ for $\epsilon_{\v k}$ far below $\mu$ o
 n the $\Delta$ scale and $1/\phi_{\v k}\simeq 2e^{1\rho_0V}(\epsilon_{\v k}-\mu)/\Omega$ for $\epsilon_{\v k}$ far above $\mu$ on this scale.

\begin{figure}[htbp]
 
\begin{center}
\includegraphics[width=0.45\textwidth]{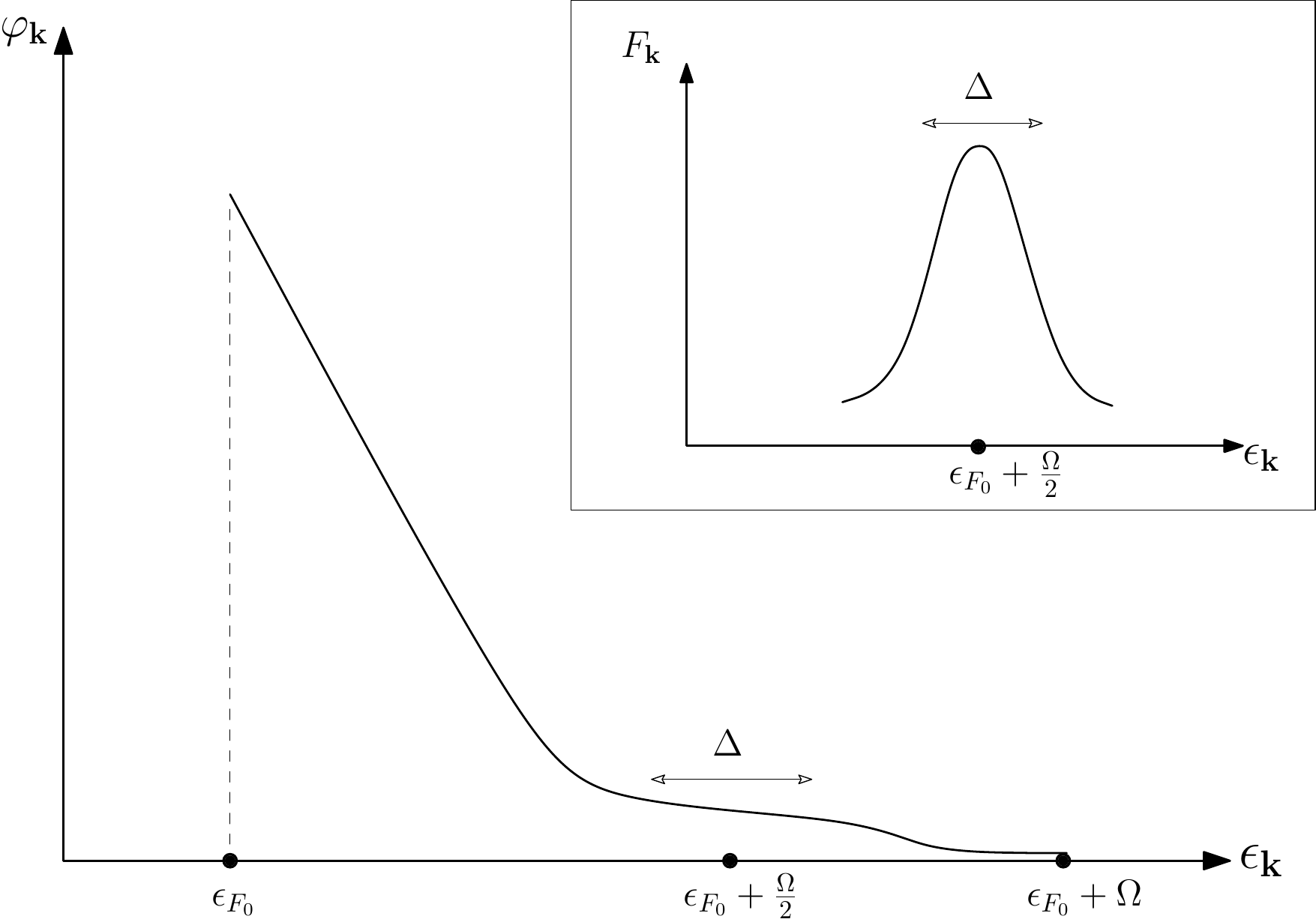}
\caption{The Cooper pair wave function $\varphi_{\v{k}}$ essentially has a linear decrease between $\epsilon_{F_0}$ and $\epsilon_{F_0}+\frac{\Omega}{2}\simeq\mu$ and a very small tail between $\mu$ and $\epsilon_{F_0}+\Omega$. \\ Insert: the ``pair wave function'' defined as $F_{\v{k}}=u_{\v{k}}v_{\v{k}}^*$ is concentrated on a $\Delta$ scale around   $\epsilon_{F_0}+\frac{\Omega}{2}$.\label{wavfig}} 

\end{center}
\end{figure}

If we now turn to the normalized distribution, i.e., the Cooper pair wave function in momentum space, it reads $\varphi_{\v k}=\phi_{\v k}/\sqrt{S}$ where, using Eq. (\ref{phi})
\begin{equation}
S=\sum_{\v k}|\phi_{\v k}^2|=N_{\Omega}(1+\frac{\Omega^2}{6\Delta^2})\simeq \frac{N_{\Omega}}{6}\,e^{2/\rho_0V},
\end{equation}
 $N_{\Omega}=\rho_0\Omega$ is the total number of pair states in the potential layer for a constant density of states $\rho_0$. Since  $e^{-1/\rho_0V}$ is very small in the small coupling limit, the Cooper pair wave function $\varphi_{\v{k}}$ is sizeable between $\epsilon_{F_0}$ and $\mu-\Delta/2$ only, where $\epsilon_{F_0}$  is the Fermi energy of the non-interacting electrons (see Fig. \ref{potfig}). $\varphi_{\v{k}}$ then  scales, within irrelevant numerical prefactors,  as 
\begin{equation}
\varphi_{\v{k}}^{(1)}\simeq \frac{1}{\sqrt{N_{\Omega}}}\frac{\mu-\epsilon_{\v k}}{\Omega}.
\end{equation}
It has a small tail (see Fig. \ref{wavfig}) of the order of
\begin{equation}
\varphi_{\v{k}}^{(2)}\simeq \frac{e^{-1/\rho_0V}}{\sqrt{N_{\Omega}}}
\end{equation}
for electron energies in the range $\pm \Delta$ around $\mu$. For higher energy, i.e., for $\epsilon_{\v k}$ between $\mu+\Delta$ and $\epsilon_{F_0}+\Omega$, the wave function is even smaller, being of the order of 
\begin{equation}
\varphi_{\v{k}}^{(3)}\simeq \frac{e^{-2/\rho_0V}}{\sqrt{N_{\Omega}}}\frac{\Omega}{\epsilon_{\v k}-\mu}.
\end{equation}

 This shows that the sizeable part of the Cooper pair wave function $\varphi_{\v k}$, which is the normalized form of $\phi_{\v k}=v_{\v k}/u_{\v k}$, is a linearly decreasing function of $\epsilon_{\v k}$ between the non-interacting electron Fermi sea $\epsilon_{F_0}$ and the normal electron Fermi sea  $\epsilon_{F}=\epsilon_{F_0}+\Omega/2$, in the case of a potential extending symmetrically on both sides of this Fermi sea. The number of pair states with sizeable weight in the Cooper pair distribution thus is of the order of $N_\Omega/2$. This understanding  has to be contrasted with what is often called "pair wave function", namely $F_{\v k}=v_{\v k}^*u_{\v k}$, and which is highly peaked at $\epsilon_{F}$ (see insert of Fig. \ref{wavfig}). $F_{\v k}$ is physically related to the excitation of electron-hole pairs in the BCS condensate while $\varphi_{\v k}$ is associated to the ground state of  up and down spin electrons added to the non-interacting Fermi sea $\epsilon_{F_0
 }$ and paired by the reduced  BCS potential.

  For completeness, in addition to Cooper pairs with creation operator $B_0^\dag$ making the BCS condensate, we must mention the ``single Cooper pair" creation operator derived by Cooper when studying a single pair of up and down spin electrons added to the $\epsilon_{F_0}$ Fermi sea. Its ( unnormalized ) creation operator reads 
  \begin{equation}
B_{N=1}^\dag=\sum_{\v k}\frac{1}{2\epsilon_{\v k}-E_1}a_{\v k}^\dag b_{-\v k}^\dag.
\end{equation}
the sum being taken over $\epsilon_{F_0}<\epsilon_{\v{k}}<\epsilon_{F_0}+\Omega$. The single pair binding energy is $E_1=2\epsilon_{F_0}-\epsilon_{c}$, where $\epsilon_{c}\simeq2\Omega e^{-2/\rho_0V}$ for small $V$. The above equation shows that the wave function of the $B_{N=1}^\dag|0\rangle$ state is concentrated on a $\epsilon_{c}$ scale above $\epsilon_{F_0}$; so, that the amount of pair states with sizeable weight in $B_{N=1}^\dag$ is of the order of $N_c=\rho_0\epsilon_{c}$ which is far smaller than the number of pairs in the $B_0^\dag$ operator making the BCS condensate.

  \section{Conclusion}

In this work, we address the BCS condensate of Cooper pairs in the canonical ensemble. To easily handle the Pauli exclusion principle between a given number of paired electrons, we first develop a formalism appropriate to Cooper pairs which is based on a set of commutators.  We then use it to, in particular, show that the Pauli exclusion principle between Cooper pairs is fully responsible for the correct value of the probability distribution peak for $N$-pair states in the BCS wave function ansatz. The standard grand canonical ensemble approach  tends to mask the key role played by Pauli blocking in this problem. We end by reconsidering what should be called ``pair wave function'' through a comparison with other composite bosons like Wannier and Frenkel excitons. 

\textbf{Acknowledgments}
We wish to thank Tony Leggett for very many discussions on the canonical ensemble approach to Cooper pairs. M. C. wants to thank the Institute of Condensed Matter Physics of the University of Illinois at Urbana-Champaign for various invitations during which most of this work has been performed.  G. Z. is supported partly by USA NSF under grant No. DMR 09-06921.


\end{document}